\documentclass[12pt]{article}

\usepackage[utf8]{inputenc}

\usepackage{authblk}

\usepackage{setspace}

\usepackage[margin=1.25in]{geometry}

\usepackage{graphicx}

\usepackage{subcaption}

\usepackage{amsmath}

\usepackage{lipsum}

\usepackage[english]{babel}

\usepackage[square,numbers]{natbib}

\begin{document}

\title{Possible formation of Quark-Gluon Plasma in small collision systems at the Large Hadron Collider: Observations and Challenges}
\author[1*]{Raghunath Sahoo}

\affil[1]{Department of Physics, Indian Institute of Technology Indore, Simrol, Khandwa Road, Indore-543552, M.P., India}
\affil[*]{Address correspondence to: Raghunath.Sahoo@cern.ch}

\onehalfspacing
\maketitle

\date{}

\begin{abstract}
With the advent of unprecedented collision energy at the Large Hadron Collider, CERN, Geneva,  a new domain of particle
production and possible formation of Quark-Gluon Plasma (QGP) in high-multiplicity proton-proton collisions and the 
collisions of light nuclei has been a much-discussed topic recently. In this review, I discuss some of the recent 
observations leading to such a possibility, associated challenges, and some predictions for the upcoming light-nuclei collisions at the LHC.

\end{abstract}

\newpage

\section{Introduction}
The creation and characterization of a deconfined state of the primordial matter of quarks and gluons, collectively called
partons, has been a frontier of high-energy nuclear physics. To create such a state of matter, which is believed to have been formed a few microseconds after the Big Bang, has taken nuclear physicists to the extremes of accelerator technology with higher energy and luminosity frontiers. Quark-Gluon Plasma (QGP) is formed in GeV-TeV collisions of heavy-nuclei, where proton-proton (pp) collisions have traditionally served as a baseline measurement to characterize heavy-ion collisions and possible formation of a QCD medium. However, the TeV pp collisions at the Large Hadron Collider (LHC) have revealed some of the QGP signatures
creating a new domain of particle production and hence questioning the validity of pp collisions to be a baseline measurement. 

In this review, we present some of the recent measurements at the LHC showing QGP-like properties in pp collisions, theoretical understanding, and challenges in hand to understand a possible formation of QGP in hadronic or light-nuclei collisions at the TeV energies.

\section{Discussion}
\begin{figure}[h!]
\includegraphics[scale=0.20]{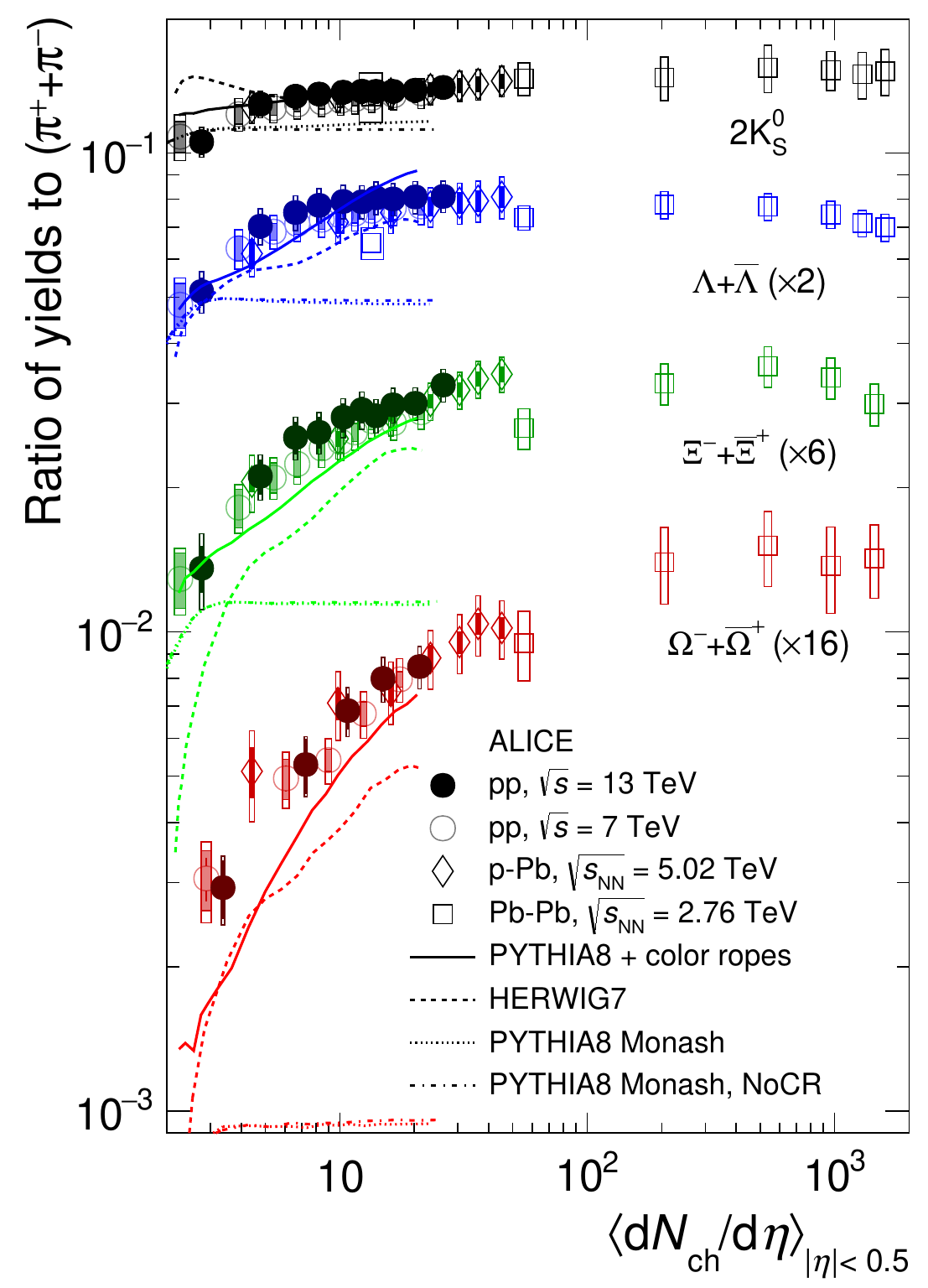}
\includegraphics[scale=0.27]{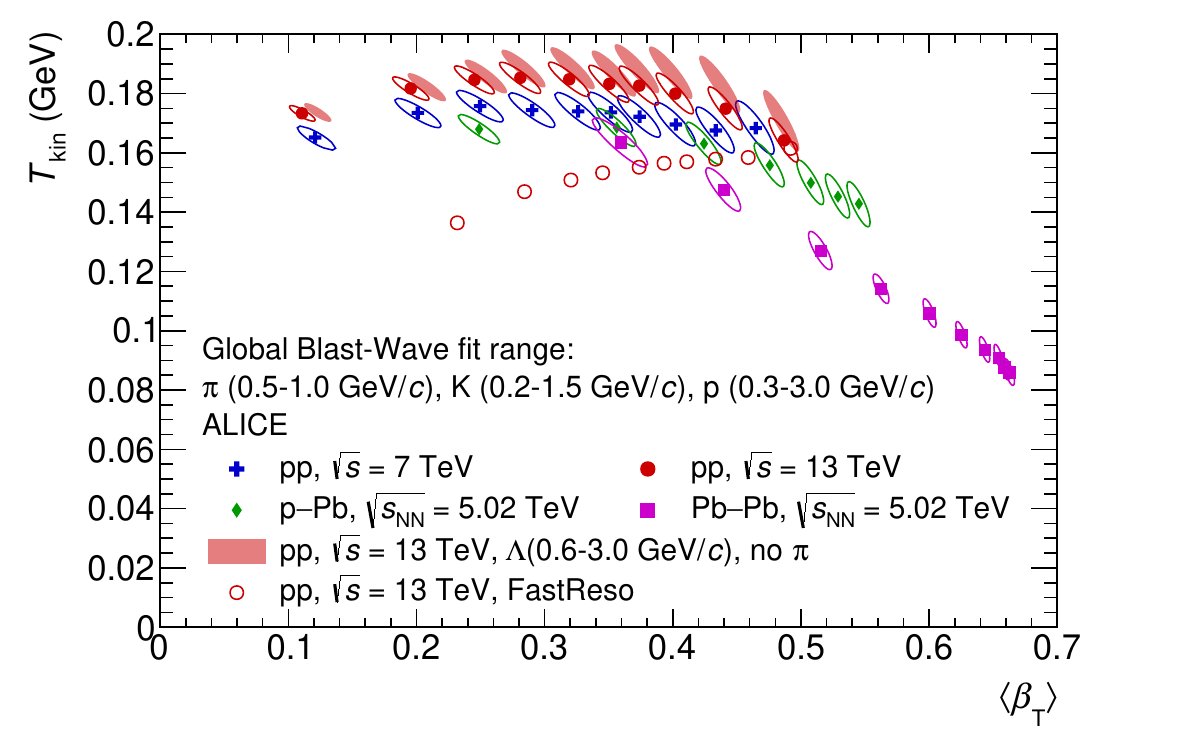}
\includegraphics[scale=0.30]{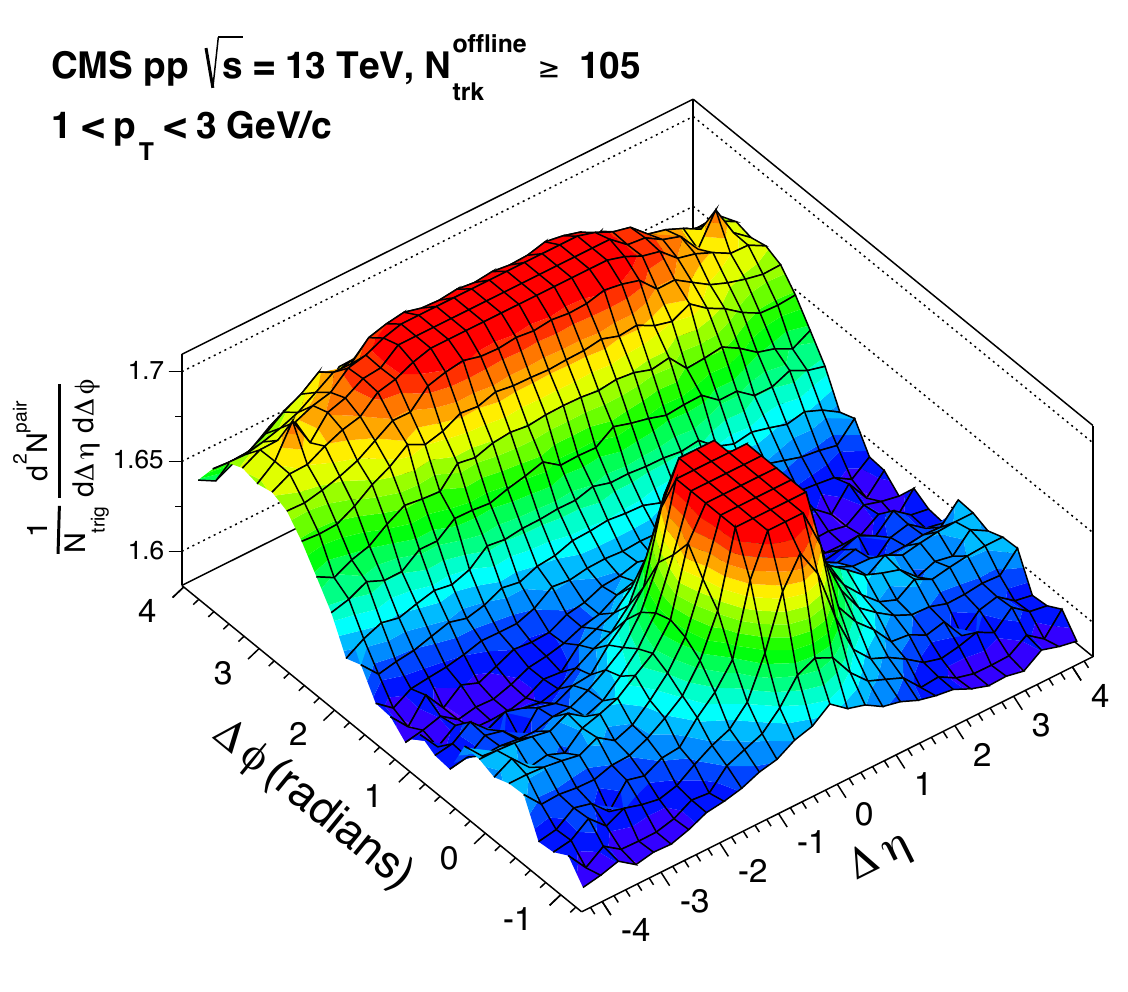}
\caption{(Left) Ratios of strange and multi-strange hadrons to pions as a
function of mid-rapidity charged particle density for pp, p–Pb and Pb–
Pb collisions at CERN LHC energies. The results show heavy ion-like
strangeness enhancement in high-multiplicity pp collisions, (Middle) Kinetic freeze-out temperature and radial flow velocity for
pp, p–Pb and Pb–Pb collisions at CERN Large Hadron Collider (LHC)
energies \cite{ALICE:2020nkc}, (Right) Two-particle correlation function in high-multiplicity pp collisions at $\sqrt{s}$ = 13 TeV for pairs of charged particles showing the
ridge structure, with each particle within $1 < p_T < 3$ GeV/c \cite{CMS:2015fgy}.}
\label{fig1}
\end{figure}

Enhancement of strangeness has been conjectured as a smoking-gun signature of QGP \cite{Rafelski:1982pu} and has been observed in heavy-ion collisions since SPS \cite{WA97:1999ikd}, RHIC \cite{STAR:2011fbd}, and then LHC \cite{ALICE:2013xmt}. Recent findings in pp collisions at $\sqrt{s}$ = 7 and 13 TeV show enhancement of strange and multistrange hadrons \cite{ALICE:2016fzo,ALICE:2020nkc}. In particular, high-multiplicity pp collisions seem to show a similar degree of strangeness enhancement as compared to heavy-ion collisions at the TeV energies (shown in Fig. \ref{fig1} (left)). This makes the TeV pp collisions more interesting. Let's now delve into other QGP-like signatures in pp collisions at the LHC. The kinetic freeze-out temperature, $T_{\rm kin}$ and collective radial flow velocity, $\beta_{\rm T}$ extracted from a simultaneous Blastwave fit to the spectra of $\pi, K$ and $p$ measured in pp collisions at $\sqrt{s}$ = 7 TeV show $T_{\rm kin} = 163 \pm 10$ MeV and $<\beta_{\rm T}> = 0.49 \pm
0.02$. This temperature is comparable with that one obtains in heavy-ion collisions and also with the quark-hadron deconfinement temperature, while the degree of collectivity is also higher  (shown in Fig. \ref{fig1} (middle)) \cite{ALICE:2020nkc}. Strong hydrodynamic collective expansion of the strongly interacting matter leads to an azimuthally collimated long-range (large $\Delta\eta$), near-side (small $\Delta\phi$) ridge-like structure in the two-particle correlations, which was seen in heavy-ion collisions \cite{STAR:2009ngv,CMS:2013bza}. Although one doesn't expect such a ridge to be formed in pp collisions, the CMS experiment 
at the LHC has reported a same-side ($\Delta\phi \sim$  0) ridge in the two-particle correlations produced in high-multiplicity pp collisions (shown in Fig. \ref{fig1} (right)) \cite{CMS:2015fgy}. These are some of the observations that make high-multiplicity pp collisions at the LHC energies very interesting and probable candidates for
possible QGP formation. A review of these findings for general readers can be found in Ref. \cite{Sahoo:2021aoy}.

Although a complete understanding of QGP-like behavior in pp collisions is not yet available both from theoretical and experimental fronts, there have been a lot of works ongoing in this direction.  The use of pp collisions as a baseline for studying the formation of a QCD medium in heavy-ion collisions has become a challenge. In view of this, recently, there are developments to consider proton
as an extended object within the framework of a Glauber model and estimate the number of participants and binary collisions.
In this approach, motivated by the shape of the structure-function obtained in deep inelastic scattering one considers the proton having an anisotropic and inhomogeneous density profile with three effective quarks connected by gluonic flux tubes. The
densities of quarks and gluons are taken as the Gaussian type assuming a spherically symmetric distribution of quark densities from their respective centers, whereas, cylindrically symmetric gluon densities about the line joining two adjacent quarks are considered. This approach has been successful in explaining the experimentally measured charged particle multiplicity and elliptic flow in pp collisions
at $\sqrt{s}$ = 13 TeV. In order to understand the possibility of a formation of a QCD medium in high-multiplicity pp collisions, a new variable is introduced, which is like the nuclear modification factor used in heavy-ion collisions. Details of this work could be found in Ref. \cite{Deb:2019yzc}. Furthermore, as one probes the interior of protons with low Bjorken-x with the advent of TeV collision energies,
more partons participate in the inelastic collisions. The concept of multi-partonic interactions (MPI) has been recently conjectured and is in use in pQCD-based event generators like PYTHIA8. This approach is successful in explaining various observations in TeV pp collisions, including the reason for having high-multiplicity in pp collisions as compared to minimum-bias pp 
collisions. One of the observations at the LHC is the production of charmonia in pp collisions \cite{Thakur:2017kpv,ALICE:2021zkd}. 
However, it should be noted here that none of the theoretical models have been successful in explaining the complete features of
charmonia production at the LHC.

\begin{figure}[h!]
\includegraphics[scale=0.30]{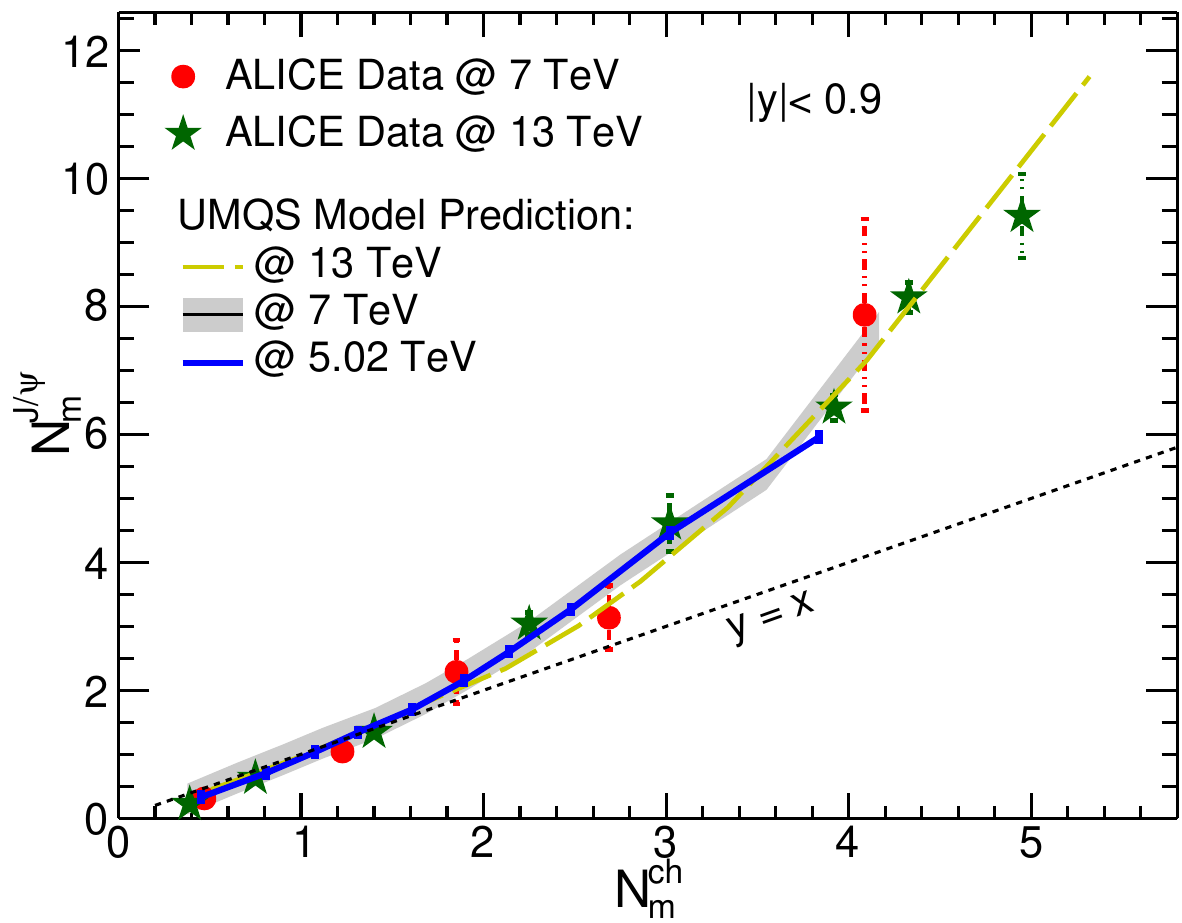}
\includegraphics[scale=0.32]{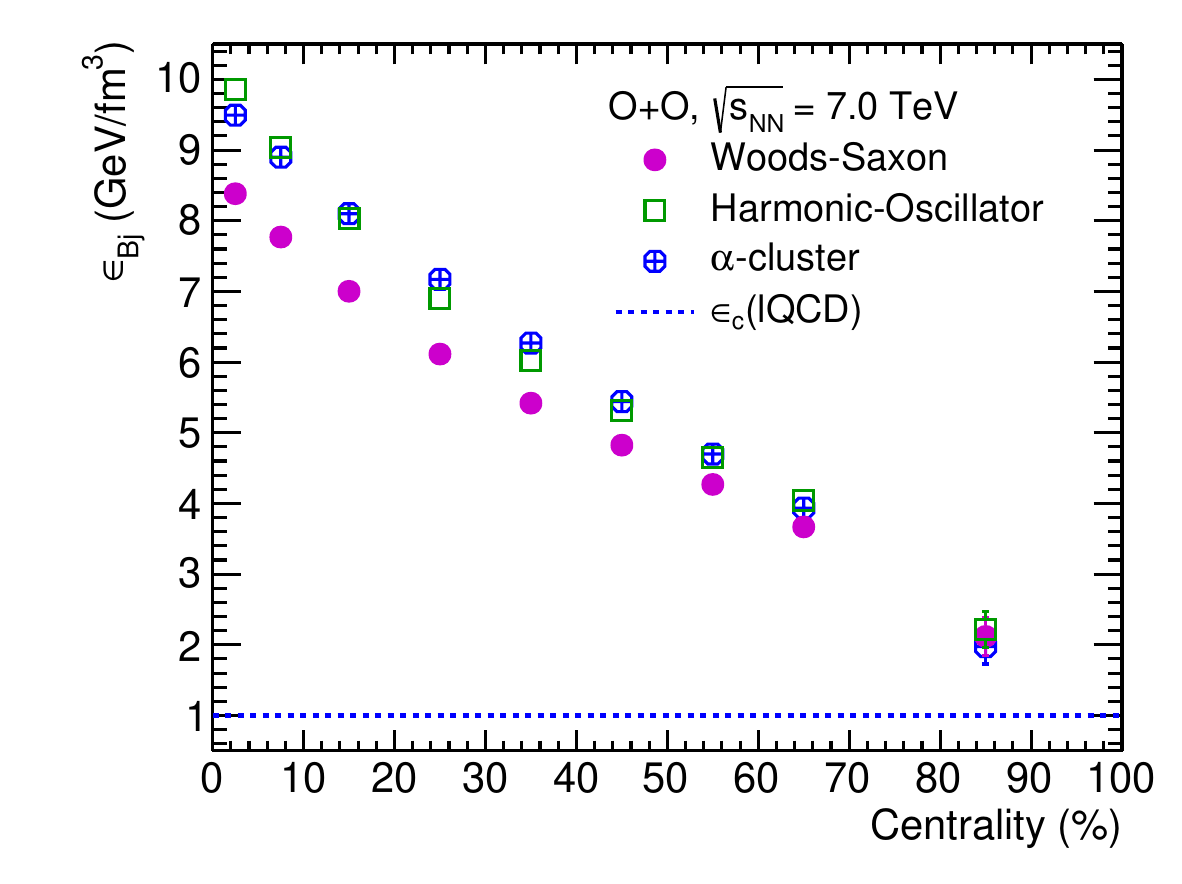}
\caption{(Left) The normalized inclusive $p_{T}$-integrated $J/\psi$ yield as a function of normalized charged-particle multiplicity at mid-rapidity is compared with experimental data corresponding to ALICE V0M selection in p$-$p collisions at $\sqrt{s}$ = 7 and 13 TeV. Model prediction for normalized $J/\psi$ production at $\sqrt{s}= 5.02$ corresponding to V0M selection in p-p collisions, is also shown here \cite{Singh:2021evv}. (Right) Bjorken energy density as a function centrality classes for pions, kaons, and protons in O-O collisions at $\sqrt{s_{\rm NN}}$ = 7 TeV. The solid (open)markers represent the Woods-Saxon (harmonic oscillator) density profile, and the markers with a cross represent the results from $\alpha$-clustered structure \cite{Behera:2021zhi}.}
\label{fig2}
\end{figure}

Going one step ahead, it would be interesting to observe the suppression of $J/\psi$ in pp collisions with a suitable baseline, 
which is considered an important signature of QGP. In the meantime, the UMQS model which incorporates the suppression of $J/\psi$ through color screening, gluonic dissociation, and collision damping, additionally, considering the regeneration of charmonium due to correlated $c-\bar{c}$ pairs, has been successful in explaining the production of normalized $J/\psi$ as a function of normalized charged particle yield \cite{Singh:2021evv}. This is shown in Fig. \ref{fig2} (left), where one defines:

\begin{equation}
N^{J/\psi}{m} = \frac{dN_{J/\psi}/d\eta}{\langle dN_{J/\psi}/d\eta \rangle}\;\;\;\;\;\;\;\;\ N^{ch}{m} = \frac{dN_{ch}/d\eta}{\langle dN_{ch}/d\eta \rangle}. \nonumber
\end{equation}

Here, $N^{J/\psi}{m}$ and $N^{ch}{m}$ are normalized with respect to their corresponding mean values in minimum bias pp collisions.

In view of these complex observations, event-shape studies are in place in order to understand the particle production dynamics
in pp collisions, by separating soft low transverse momentum transfer processes from the pQCD-based hard scattering ones. In this
direction, event-shape observables like sphericity \cite{ALEPH:2003obs}, spherocity \cite{Cuautle:2014yda,Deb:2020ige,Prasad:2021bdq,Mallick:2021hcs}, $R_{\rm T}$ \cite{Martin:2016igp,ALICE:2023yuk}, and flatenicity \cite{Ortiz:2022mfv} are of worth mentioning.

Light-ion collisions at the LHC energies would be highly interesting while bridging the multiplicity gap between hadronic and
heavy-ion collisions. In this direction, proton-Oxygen (p-O) and Oxygen-Oxygen (O-O) collisions are a way forward in order to understand 
the high energy cosmic ray interactions and also QGP formation in small systems. Recently, there have been several theoretical and phenomenological works to understand particle production and signatures of QGP in O-O collisions \cite{Behera:2023nwj,Behera:2021zhi,Li:2020vrg,Rybczynski:2019adt}. Most interesting is to study
the effect of the nuclear density profile of the Oxygen nucleus on various observables and make a prediction of the one which can
reveal the $\alpha$-clustering density profile, which is seen in low-energy nuclear physics experiments \cite{Wheeler:1937zza,Kanungo:2009zz}. It is seen that across all centrality O-O collisions, one produces initial Bjorken energy densities
which are higher than the lattice QCD predicted value of around $\rm1 ~GeV/fm^3$ required for a deconfinement transition (shown in Fig. \ref{fig2} (right))\cite{Behera:2021zhi}. This makes O-O collisions more interesting as a case of light-nuclei collisions and possible formation
of QGP.

\section{Conclusion}
In view of the complex and emerging phenomena in particle production dynamics in TeV pp collisions, along with the observations of QGP-like signatures, pp collisions in general, are no longer suitable for studying the formation of QCD medium in heavy-ion collisions
considering them as a baseline. 
High-multiplicity pp collision events are of utmost importance to understanding the underlying production dynamics and possible formation of QGP-droplets. The small systems are the test-bed in looking for a threshold to study thermalization, and applicability of hydrodynamics. 

However, in the absence of QGP signatures like jet-quenching, ${J/\psi}$ suppression in pp collisions at the LHC energies, the search for QGP-droplets in small collision systems stays elusive. The proposed proton-Oxygen and Oxygen-Oxygen collisions at the LHC
energies may address some of the issues in these directions. In view of these, high-energy hadronic and light-ion collisions at the
TeV energies with high luminosity will be an exciting domain, which will enhance our understanding of QGP physics while extending the
measurements to the heavy-flavor sector with low transverse momentum reach and higher statistics.
\bibliography{bibitems}


\end{document}